\documentclass[pra,twocolumn,superscriptaddress,amsmath,showpacs]{revtex4}
\usepackage{mathrsfs}
\usepackage{graphicx}
\usepackage{pifont}
\usepackage{color}

\newcommand{\bra}[1]{\langle #1 |}
\newcommand{\ket}[1]{| #1 \rangle}

\newcommand{\ketbra}[2]{| #1 \rangle \langle #2 |}
\newcommand{\expect}[1]{\langle #1 \rangle}

\begin{document}
\title{Quantifying non-Gaussianity of a quantum state by the negative entropy of quadrature distributions}
\author{Jiyong Park}
\email{jiyong.park@hanbat.ac.kr}
\affiliation{School of Basic Sciences, Hanbat National University, Daejeon, 34158, Korea}
\author{Jaehak Lee}
\affiliation{School of Computational Sciences, Korea Institute for Advanced Study, Seoul, 02455, Korea}
\author{Kyunghyun Baek}
\affiliation{School of Computational Sciences, Korea Institute for Advanced Study, Seoul, 02455, Korea}
\author{Hyunchul Nha}
\affiliation{Department of Physics, Texas A\&M University at Qatar, Education City, P.O. Box 23874, Doha, Qatar}
\date{\today}

\begin{abstract}
We propose a non-Gaussianity measure of a multimode quantum state based on the negentropy of quadrature distributions. Our measure satisfies desirable properties as a non-Gaussianity measure, i.e., faithfulness, invariance under Gaussian unitary operations, and monotonicity under Gaussian channels. Furthermore, we find a quantitative relation between our measure and the previously proposed non-Gaussianity measures defined via quantum relative entropy and the quantum Hilbert-Schmidt distance. This allows us to estimate the non-Gaussianity measures readily by homodyne detection, which would otherwise require a full quantum-state tomography.
\end{abstract}

\maketitle

\section{Introduction}
In continuous-variable (CV) quantum information \cite{Braunstein2005, Weedbrook2012}, non-Gaussian resources are essential as there exist quantum information tasks not achievable by Gaussian counterparts only. For example, it is impossible to distill entanglement of Gaussian states by Gaussian operations and measurements \cite{Eisert2002, Fiurasek2002, Giedke2002, Kraus2003}, to manifest nonlocality of Gaussian states with Gaussian measurements \cite{Nha2004, Garcia2004}, and to correct quantum errors in Gaussian quantum communication protocols \cite{Niset2009}. Furthermore, non-Gaussian resources can be used to provide substantial advantages over Gaussian counterparts. Non-Gaussian entanglement can survive under Gaussian noises longer than Gaussian entanglement \cite{Allegra2010, Adesso2011, Sabapathy2011, JL2011, Allegra2011}. Non-Gaussian operations can enhance the nonclassical properties, e.g., optical nonclassicality \cite{Agarwal1990, SYL2010}, quantum entanglement \cite{Takahashi2010, SYL2011, Navarrete2012, SYL2013, JL2013, Kurochkin2014, Ulanov2015, Hu2017}, nonlocality \cite{Park2012}, as well as the performance in quantum information protocols including quantum teleportation \cite{Opatrny2000, Cochrane2002, Olivares2003, DellAnno2007, JL2017}, quantum dense coding \cite{Kitagawa2005}, quantum linear amplification \cite{Nha2010, Zavatta2011, Kim2012}, quantum key distribution \cite{Huang2013}, and quantum illumination \cite{Fan2018}.

For the purpose of addressing the role of non-Gaussianity in CV quantum information rigorously, it is desirable to quantify the non-Gaussianity of quantum resources. There have been several proposals to characterize the {\it non-Gaussianity} of quantum states, e.g., by means of quantum Hilbert-Schmidt distance \cite{Genoni2007, Genoni2010}, quantum relative entropy \cite{Genoni2008, Genoni2010}, Wehrl entropy \cite{Ivan2012}, quantum R{\'e}nyi relative entropy \cite{Baek2018}, Wigner-Yanase skew information \cite{Fu2020}, and trace overlap \cite{Mandilara2010} (see Table I). On the other hand, {\it quantum non-Gaussianity} has been introduced as a stronger form of non-Gaussianity considering a convex mixture of Gaussian states as a free state in quantum resource theory. It has been quantified by employing negative volume in phase space \cite{Albarelli2018, Takagi2018}, quantum relative entropy \cite{Park2019}, the number of zeros in Husimi-Q function \cite{Chabaud2020}, and robustness \cite{Lami2021} (see Table II). We here focus on non-Gaussianity measures that quantify the deviation of a quantum state from its reference Gaussian state. While the previously proposed measures have provided a useful basis for analyzing non-Gaussian resources, it is difficult to determine the values of those measures without full information on the state under examination. With this in mind, for the case of non-Gaussianity measure via quantum relative entropy, an observable lower bound was provided using the statistics from a photon-number-resolving detector in \cite{Genoni2010}. However, it works when there is \textit{a priori} information, i.e., the covariance matrix of a quantum state. It is thus natural to ask whether the non-Gaussianity of a quantum state can be estimated by a readily accessible measurement setup, e.g., homodyne detection. Qualitatively speaking, if a quadrature distribution of a quantum state measured by homodyne detection has a non-Gaussian profile, it is evident that the quantum state is non-Gaussian. But it is worth investigating in what rigorous context the non-Gaussianity manifested by a quadrature distribution can be adopted to define a desirable non-Gaussianity measure for a general multimode quantum state.

Here we propose a non-Gaussianity measure of a quantum state in terms of the maximum negentropy of quadrature distributions. Our measure fulfills several desirable properties, i.e., non-negativity, faithfulness, invariance under Gaussian unitary operations, nonincreasing under Gaussian channels. Furthermore, we show that our measure provides lower bounds for the non-Gaussianity measures based on quantum relative entropy \cite{Genoni2008} and quantum Hilbert-Schmidt distance \cite{Genoni2007}, respectively. These quantitative connections make our approach useful to address a general quantum non-Gaussian state by a highly efficient homodyne detection.

    \begin{table*}
    \begin{ruledtabular}
    \begin{tabular}{lcccc}
        Measure & Faithfulness & $\mathcal{N} ( \hat{U}_{\mathrm{G}} \rho \hat{U}_{\mathrm{G}}^{\dag} ) = \mathcal{N} ( \rho )$ & $\mathcal{N} ( \mathcal{T}_{\mathrm{G}} [ \rho ] ) \leq \mathcal{N} ( \rho )$ & Observable lower bound \\
        \hline
        Quantum Hilbert-Schmidt distance \cite{Genoni2007} & \ding{51} & \ding{51} & ? & This paper \\
        \hline
        Quantum relative entropy \cite{Genoni2008} & \ding{51} & \ding{51} & \ding{51} & This paper and Ref. \cite{Genoni2010} \\
        \hline
        Wehrl entropy \cite{Ivan2012} & \ding{51} & \ding{55} & \ding{55} & \\
        \hline
        Quantum R{\' e}nyi relative entropy \cite{Baek2018} & \ding{51} & \ding{51} & \ding{51} & This paper for $\alpha \geq 1$ \\
        \hline
        Wigner-Yanase skew information \cite{Fu2020} & \ding{51} & \ding{55} & \ding{55} & \\ \hline
        Trace overlap \cite{Mandilara2010} & \ding{55} & \ding{51} & \ding{55} & This paper for upper bound \\ \hline
        Kullback-Leibler divergence & \ding{51} & \ding{51} & \ding{51} & This paper
    \end{tabular}
    \caption{Comparison table for the non-Gaussianity measures. A non-Gaussianity measure $\mathcal{N} ( \rho )$ satisfies the faithfulness condition when the measure becomes zero if and only if the state $\rho$ is Gaussian. $\mathcal{N} ( \hat{U}_{\mathrm{G}} \rho \hat{U}_{\mathrm{G}}^{\dag} ) = \mathcal{N} ( \rho )$ and $\mathcal{N} ( \mathcal{T}_{\mathrm{G}} [ \rho ] ) \leq \mathcal{N} ( \rho )$ indicate that the non-Gaussianity measure $\mathcal{N} ( \rho )$ is invariant under a Gaussian unitary operation $\hat{U}_{\mathrm{G}}$ and nonincreasing under a Gaussian channel $\mathcal{T}_{\mathrm{G}}$, respectively.}
    \end{ruledtabular}
    \end{table*}

\section{Non-Gaussianity measure by classical relative entropy}
In classical information theory, a representative measure for the non-Gaussianity of a probability distribution is negentropy \cite{Schrodinger1944}. It quantifies the relative entropy between a given probability distribution $X$ and its reference Gaussian distribution $X_{\mathrm{G}}$ having the same mean and variance as $X$,
	\begin{equation} \label{eq:J}
		J ( X ) \equiv D_{\mathrm{KL}} ( X || X_{\mathrm{G}} ),
	\end{equation}
 where $D_{\mathrm{KL}} ( X || Y ) = \int d\mu X (\mu) [ \ln X (\mu) - \ln Y (\mu) ]$ is the Kullback-Leibler divergence \cite{Erven2014}, also known as classical relative entropy, between two probability distributions $X$ and $Y$. It is known that Eq.~\eqref{eq:J} can be rewritten simply as
	\begin{equation}
		J ( X ) = H ( X_{\mathrm{G}} ) - H ( X ),
	\end{equation}
where $H ( X ) = - \int d\mu X ( \mu ) \ln X ( \mu )$ is the differential entropy of a probability distribution $X$ \cite{CoverThomas}.

\subsection{Our measure}
We here define a non-Gaussianity measure of an $N$-mode quantum state $\rho$ by means of negentropy as
	\begin{equation} \label{eq:NGKL1}
		\mathcal{N}_{\mathrm{KL}} ( \rho ) = \max_{\Theta, \Phi} J_{\rho} ( Q_{\Theta, \Phi} ),
	\end{equation}
where $Q_{\Theta, \Phi}$ denotes a probability distribution for an $N$-mode quadrature operator $\hat{Q}_{\Theta, \Phi}$ given by
    \begin{equation} \label{eq:NMQO}
        \hat{Q}_{\Theta, \Phi} = \sum_{j=1}^{N} c_{j} \hat{q}_{j, \phi_{j}}.
    \end{equation}
Here $\hat{q}_{j, \phi_{j}} = \frac{1}{\sqrt{2}} ( \hat{a}_{j} e^{i\phi_{j}} + \hat{a}_{j}^{\dag} e^{-i\phi_{j}} )$ is a quadrature amplitude for the $j$th mode, $\Phi = ( \phi_{1}, \phi_{2}, ..., \phi_{N} )^{T}$ the set of quadrature phases $\phi_{j}$, and $\Theta = ( \theta_{1}, \theta_{2}, ..., \theta_{N-1} )^{T}$ the set of angular coordinates that determines the superposition coefficient $c_{j}$ in Eq. (4) as
    \begin{equation}
        c_{j} = \begin{cases}
                    \cos \theta_{1} & \mbox{for $j = 1$,} \\
                    \cos \theta_{j} \prod_{k=1}^{j-1} \sin \theta_{k} & \mbox{for $1 < j < N$,} \\
                    \prod_{k=1}^{N-1} \sin \theta_{k} & \mbox{for $j = N$.}
                \end{cases}
    \end{equation}
Before introducing the properties of our measure, we briefly explain how the probability distribution $Q_{\Theta, \Phi}$ can be experimentally accessible. Using a Heisenberg picture, we see that the $N$-mode quadrature $\hat{Q}_{\Theta, \Phi}$ in Eq.~\eqref{eq:NMQO} can be addressed via a linear optical network composed of beam splitters and phase shifters as
    \begin{equation} \label{eq:HP}
        \hat{Q}_{\Theta, \Phi} = \hat{L}^{\dag} \hat{q}_{1, 0} \hat{L},
    \end{equation}
where the Gaussian unitary operation $\hat{L}$ for the linear optical network is given by
    \begin{equation}
        \hat{L} = \hat{B}_{1,2} ( \theta_{1} ) \cdots \hat{B}_{N-1,N} ( \theta_{N-1} ) \hat{R}_{1} ( \phi_{1} ) \cdots \hat{R}_{N} ( \phi_{N} ),
    \end{equation}
with $\hat{R}_{j} ( \phi ) = \exp ( i \phi \hat{a}_{j}^{\dag} \hat{a}_{j} )$ and $\hat{B}_{j,k} ( \theta ) = \exp ( \theta \hat{a}_{j}^{\dag} \hat{a}_{k} -  \theta \hat{a}_{k}^{\dag} \hat{a}_{j} )$ representing the phase rotation on the $j$th mode and the beam-splitting operation between the $j$th and $k$th modes with the transmittance $T = \cos^{2} \theta$, respectively (see Fig. 1). Using $\hat{R}_{j}^{\dag} ( \phi ) \hat{q}_{j, 0} \hat{R}_{j} ( \phi ) = \hat{q}_{j, \phi}$ and $\hat{B}_{jk}^{\dag} ( \theta ) q_{j, 0} \hat{B}_{jk} ( \theta ) = \cos \theta \hat{q}_{j, 0} + \sin \theta \hat{q}_{k, 0}$ \cite{Olivares2012, BarnettRadmore}, Eq.~\eqref{eq:HP} gives the result in Eq. (4). The relation in Eq.~\eqref{eq:HP} implies that we obtain the probability distribution $Q_{\Theta, \Phi}$ by a single-mode homodyne detection using a linear optical network. Note that one can fully reconstruct the $N$-mode quantum state $\rho$ by examining the whole set of $Q_{\Theta, \Phi}$ \cite{D'Ariano1999}.

    \begin{table*}
    \begin{ruledtabular}
    \begin{tabular}{lcccc}
        Measure & Faithfulness & $\mathcal{Q} ( \hat{U}_{\mathrm{G}} \rho \hat{U}_{\mathrm{G}}^{\dag} ) = \mathcal{Q} ( \rho )$ & $\mathcal{Q} ( \mathcal{T}_{\mathrm{G}} [ \rho ] ) \leq \mathcal{Q} ( \rho )$ & Observable lower bound \\
        \hline
        Negative volume in phase space \cite{Albarelli2018, Takagi2018} & \ding{55} & \ding{51} & \ding{51} & \\
        \hline
        Quantum relative entropy \cite{Park2019} & \ding{51} & \ding{51} & \ding{51} & \\
        \hline
        Number of zeros in Husimi-Q function \cite{Chabaud2020} & \ding{51} & \ding{51} & ? & \\
        \hline
        Robustness \cite{Lami2021} & \ding{51} & \ding{51} & \ding{51} & \\ \hline      
    \end{tabular}
    \caption{Comparison table for the quantum non-Gaussianity measures. A quantum non-Gaussianity measure $\mathcal{Q} ( \rho )$ satisfies the faithfulness condition when the measure becomes zero if and only if the state $\rho$ is a probabilistic mixture of Gaussian states. $\mathcal{Q} ( \hat{U}_{\mathrm{G}} \rho \hat{U}_{\mathrm{G}}^{\dag} ) = \mathcal{N} ( \rho )$ and $\mathcal{N} ( \mathcal{T}_{\mathrm{G}} [ \rho ] ) \leq \mathcal{N} ( \rho )$ indicate that the quantum non-Gaussianity measure $\mathcal{Q} ( \rho )$ is invariant under a Gaussian unitary operation $\hat{U}_{\mathrm{G}}$ and nonincreasing under a Gaussian channel $\mathcal{T}_{\mathrm{G}}$, respectively. Making a distinction between the measures in Tables I and II, we refer to a measure as a non-Gaussianity and quantum non-Gaussianity measure if it deals with the deviation of a quantum state from its reference Gaussian state and a convex set of Gaussian states, respectively.}
    \end{ruledtabular}
    \end{table*}

\subsection{Properties}

Our measure has the following properties:

1. The measure is non-negative. $\mathcal{N}_{\mathrm{KL}} ( \rho ) \geq 0$.

\textit{Proof.} It is a direct consequence of the fact that the Kullback-Leibler divergence is non-negative \cite{CoverThomas}.

2. The measure is faithful. That is, $\mathcal{N}_{\mathrm{KL}} ( \rho )$ is zero if and only if the state $\rho$ is Gaussian.

\textit{Proof.} An $N$-mode Gaussian state $\sigma$ is uniquely determined only by local position and momentum averages for each mode, i.e., $\expect{\hat{x}_{j}}_{\sigma} \equiv \expect{\hat{q}_{j, 0}}_{\sigma}$ and $\expect{\hat{p}_{j}}_{\sigma} \equiv \expect{\hat{q}_{j, \pi / 2}}_{\sigma}$, respectively, for $j \in \{ 1, ..., N \}$ with $\expect{\hat{o}}_{\rho} = \mathrm{tr} [ \rho \hat{o} ]$, and the $2N \times 2N$ covariance matrix $\Gamma ( \sigma )$ whose elements are given by $\Gamma_{jk} ( \sigma ) = \expect{\hat{Q}_{j} \hat{Q}_{k}}_{\sigma} - \expect{\hat{Q}_{j}}_{\sigma} \expect{\hat{Q}_{k}}_{\sigma}$ with $\hat{Q} = ( \hat{x}_{1}, \hat{p}_{1}, ..., \hat{x}_{N}, \hat{p}_{N} )^{T}$ \cite{Weedbrook2012}.

If the state $\rho$ is Gaussian, $\mathcal{N}_{\mathrm{KL}} ( \rho ) = 0$, as the probability distributions are Gaussian for all quadrature amplitudes $\hat{Q}_{\Theta, \Phi}$. Its converse is also true. We may introduce the reference Gaussian state $\rho_{\mathrm{G}}$ having the same means and covariance with the given state $\rho$, i.e., $\expect{\hat{Q}}_{\rho} = \expect{\hat{Q}}_{\rho_{\mathrm{G}}}$ and $\Gamma ( \rho ) = \Gamma ( \rho_{\mathrm{G}} )$. Note that the reference Gaussian state provides the reference Gaussian probability distribution for all quadratures $\hat{Q}_{\Theta, \Phi}$ as $\expect{\hat{Q}_{\Theta, \Phi}}_{\rho_{\mathrm{G}}} = \expect{\hat{Q}_{\Theta, \Phi}}_{\rho}$ and $\expect{\hat{Q}_{\Theta, \Phi}^{2}}_{\rho} = \expect{\hat{Q}_{\Theta, \Phi}^{2}}_{\rho_{\mathrm{G}}}$ always hold. In this respect we obtain an alternative expression for Eq. \eqref{eq:NGKL1} as
    \begin{equation} \label{eq:NGKL2}
        \mathcal{N}_{\mathrm{KL}} ( \rho ) = \max_{\Theta, \Phi} D_{\mathrm{KL}} ( Q_{\Theta, \Phi, \rho} || Q_{\Theta, \Phi, \rho_{\mathrm{G}}} ).
    \end{equation}
If $\mathcal{N}_{\mathrm{KL}} ( \rho ) = 0$, the state $\rho$ and its reference Gaussian state $\rho_{\mathrm{G}}$ have the identical probability distributions for all $Q_{\Theta, \Phi}$. Because the whole set of $Q_{\Theta, \Phi}$ determines a quantum state \cite{D'Ariano1999}, $\rho = \rho_{\mathrm{G}}$, i.e., the state $\rho$ must be Gaussian.

3. The measure is invariant under Gaussian unitary operations, i.e., $\mathcal{N}_{\mathrm{KL}} ( \hat{U}_{\mathrm{G}} \rho \hat{U}_{\mathrm{G}}^{\dag} ) = \mathcal{N}_{\mathrm{KL}} ( \rho )$, where $\hat{U}_{\mathrm{G}}$ is a Gaussian unitary operation.

\textit{Proof.} In the Heisenberg picture, a Gaussian unitary operation yields an affine transformation of quadrature operators \cite{Weedbrook2012} as
    \begin{equation}
        \hat{U}_{\mathrm{G}}^{\dag} \hat{Q} \hat{U}_{\mathrm{G}} = \mathsf{A} \hat{Q} + \mathsf{B},
    \end{equation}
where $\mathsf{A}$ and $\mathsf{B}$ correspond to linear transformation and translation, respectively. It indicates that a Gaussian unitary operation on an $N$-mode quadrature operator gives
    \begin{equation} \label{eq:LT}
        \hat{U}_{\mathrm{G}}^{\dag} \hat{Q}_{\Theta, \Phi} \hat{U}_{\mathrm{G}} = a \hat{Q}_{\Theta^{\prime}, \Phi^{\prime}} + b,
    \end{equation}
where $a$ and $b$ are scaling factor and translation, respectively. It is important to note that the sets of $\{ \Theta, \Phi \}$ and $\{ \Theta^{\prime}, \Phi^{\prime} \}$ have one-to-one correspondence, as the inverse of the Gaussian unitary operation $\hat{U}_{\mathrm{G}}$ is another Gaussian unitary operation, i.e., $(\hat{U}_{\mathrm{G}})^{-1}=\hat{U}_{\mathrm{G}}^{\dag}$.

A key for the proof is that the negentropy is invariant under scaling and translation, i.e., $\mathcal{L} [ X ( \mu )]  = a X ( a \mu + b )$ gives
	\begin{equation} \label{eq:INVJ}
		J ( \mathcal{L} [ X ] ) = J ( X ),
	\end{equation}
which can be readily shown as
	\begin{align}
		D_{\mathrm{KL}} ( \mathcal{L} [ X ] || \mathcal{L} [ X ]_{\mathrm{G}} ) = & D_{\mathrm{KL}} ( \mathcal{L} [ X ] || \mathcal{L} [ X_{\mathrm{G}} ] ) \nonumber \\
		= & \int d\mu \mathcal{L} [ X ( \mu )]  \ln \frac{\mathcal{L} [ X ( \mu )] }{\mathcal{L} [ X_{\mathrm{G}}( \mu ) ] } \nonumber \\
		= & \int d\mu a X ( a \mu + b ) \ln \frac{X ( a \mu + b )}{X_{\mathrm{G}} ( a \mu + b )} \nonumber \\
		= & \int d\mu^{\prime} X ( \mu^{\prime} ) \ln \frac{X ( \mu^{\prime} )}{X_{\mathrm{G}} ( \mu^{\prime} )} \nonumber \\
		= & D_{\mathrm{KL}} ( X || X_{\mathrm{G}} ),
	\end{align}
where we have used the fact that the reference Gaussian distribution of $\mathcal{L} [ X ]$ becomes $\mathcal{L} [ X_{\mathrm{G}} ]$, due to the linearity of the transformation $\mathcal{L}$.

Equipped with Eqs. \eqref{eq:LT} and \eqref{eq:INVJ}, we have
    \begin{align}
        \mathcal{N}_{\mathrm{KL}} ( \hat{U}_{\mathrm{G}} \rho \hat{U}_{\mathrm{G}} ) & =  \max_{\Theta, \Phi} J_{\hat{U}_{\mathrm{G}} \rho \hat{U}_{\mathrm{G}}^{\dag}} ( Q_{\Theta, \Phi} ) \nonumber \\
        & = \max_{\Theta^{\prime}, \Phi^{\prime}} J_{\rho} ( Q_{\Theta^{\prime}, \Phi^{\prime}} ) \nonumber \\
        & = \mathcal{N}_{\mathrm{KL}} ( \rho ).
    \end{align}

	\begin{figure}[!t]
		\includegraphics[scale=0.8]{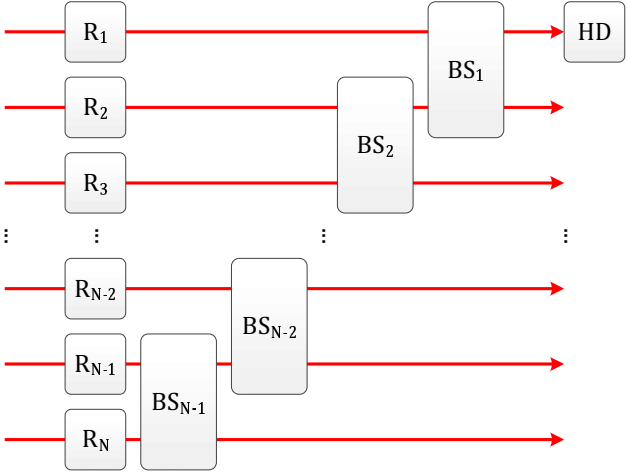}
		\caption{Linear optical network for measuring the probability distribution $Q_{\Theta, \Phi}$. $\mathrm{R}_{j}$ and $\mathrm{BS}_{k}$ represent the phase rotation on the $j$th mode and the beam-splitting operation between the $k$th and $(k+1)$th modes, respectively. Applying these Gaussian unitary operations and performing homodyne detection (HD) on the first mode, we obtain the probability distribution $Q_{\Theta, \Phi}$ for the input state.}
	    \label{fig:LON}
	\end{figure}

4. The measure is nonincreasing under partial trace. $\mathcal{N}_{\mathrm{KL}} ( \rho_{AB} ) \geq \mathcal{N}_{\mathrm{KL}} ( \rho_{A} )$.

\textit{Proof.} It is a direct consequence of the fact that the set of $\hat{Q}_{\Theta, \Phi}$ for $\rho_{AB}$ contains the one for $\rho_{A}$. 

5. The measure is invariant under the addition of Gaussian ancilla, i.e., $\mathcal{N}_{\mathrm{KL}} ( \rho \otimes \sigma ) = \mathcal{N}_{\mathrm{KL}} ( \rho )$ with a Gaussian state $\sigma$.

\textit{Proof.} We can recast a multimode quadrature operator $\hat{Q}_{\Theta, \Phi}$ for $\rho \otimes \sigma$ in the form of $\sqrt{\eta} \hat{X}_{\rho} + \sqrt{1-\eta} \hat{X}_{\sigma}$. Thus the quadrature distributions to consider for $\rho \otimes \sigma$ and its reference Gaussian state $\rho_{\mathrm{G}} \otimes \sigma$ are given by
	\begin{align}
		X_{\rho \otimes \sigma} ( x ) = & \int \frac{dy}{\sqrt{\eta (1-\eta)}} X_{\rho} ( \frac{y}{\sqrt{\eta}} ) X_{\sigma} ( \frac{x-y}{\sqrt{1-\eta}} ), \nonumber \\
		X_{\rho_{\mathrm{G}} \otimes \sigma} ( x ) = &  \int \frac{dy}{\sqrt{\eta (1-\eta)}} X_{\rho_{\mathrm{G}}} ( \frac{y}{\sqrt{\eta}} ) X_{\sigma} ( \frac{x-y}{\sqrt{1-\eta}} ),
	\end{align}
respectively. Using the data processing inequality for the Kullback-Leibler divergence \cite{Erven2014}, we have
	\begin{equation}
		D_{\mathrm{KL}} ( X_{\rho \otimes \sigma} | X_{\rho_{\mathrm{G}} \otimes \sigma} ) \leq D_{\mathrm{KL}} ( X_{\rho} | X_{\rho_{\mathrm{G}}} ),
	\end{equation}
which yields
    \begin{align}
        \mathcal{N}_{\mathrm{KL}} ( \rho \otimes \sigma ) & = \max_{\Theta, \Phi} J_{\mathrm{\rho \otimes \sigma}} ( Q_{\Theta, \Phi} ) \nonumber \\
        & \leq \max_{\Theta, \Phi} J_{\mathrm{\rho}} ( Q_{\Theta, \Phi} ) \nonumber \\
        & = \mathcal{N}_{\mathrm{KL}} ( \rho ).
    \end{align}
Combining it with the property 4, i.e., $\mathcal{N}_{\mathrm{KL}} ( \rho \otimes \sigma ) \geq \mathcal{N}_{\mathrm{KL}} ( \rho )$, we finally obtain $\mathcal{N}_{\mathrm{KL}} ( \rho \otimes \sigma ) = \mathcal{N}_{\mathrm{KL}} ( \rho )$.

6. The measure is nonincreasing under a Gaussian channel $\mathcal{T}_{\mathrm{G}}$, i.e., $\mathcal{N}_{\mathrm{KL}} ( \mathcal{T}_{\mathrm{G}} [ \rho ] ) \leq \mathcal{N}_{\mathrm{KL}} ( \rho )$.

\textit{Proof.} The action of a Gaussian channel $\mathcal{T}_{\mathrm{G}}$ is generally described by a Gaussian unitary interaction between a system and a Gaussian environment $E$, i.e., $\mathcal{T}_{\mathrm{G}} [ \rho ] = \mathrm{tr}_{\mathrm{E}} [ \hat{U}_{\mathrm{G}} ( \rho \otimes \sigma ) \hat{U}_{\mathrm{G}}^{\dag} ]$, with $\sigma$ the state of the environment \cite{Weedbrook2012}. Combining properties 3, 4, and 5, we have
	\begin{align}
		\mathcal{N}_{\mathrm{KL}} ( \mathcal{T}_{\mathrm{G}} [ \rho ] ) & \leq \mathcal{N}_{\mathrm{KL}} ( \hat{U}_{\mathrm{G}} [ \rho \otimes \sigma ] \hat{U}_{\mathrm{G}}^{\dag} ) \nonumber \\
		& = \mathcal{N}_{\mathrm{KL}} ( \rho \otimes \sigma ) \nonumber \\
		& = \mathcal{N}_{\mathrm{KL}} ( \rho ).
	\end{align}

	\begin{figure}[!t]
		\includegraphics[scale=0.8]{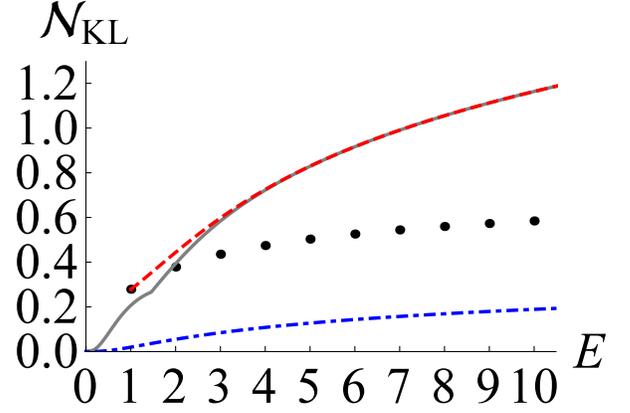}
		\caption{Our non-Gaussianity measure $\mathcal{N}_{\mathrm{KL}} ( \rho )$ for Fock states (black dots), even cat states (gray solid line), odd cat states (red dashed line), and phase-averaged coherent state (blue dot-dashed line) against the mean photon number $E = \mathrm{tr} ( \rho \hat{a}^{\dag} \hat{a} )$ for each state.}
        \label{fig:NKL1}
	\end{figure}

\subsection{Examples}
In this section, we investigate the non-Gaussianity measure for some single-mode and two-mode non-Gaussian states as examples. 
\subsubsection{Fock states}
The quadrature distribution of a Fock state $\ketbra{n}{n}$ is given by
    \begin{equation}
        P_{\ketbra{n}{n}} ( x_{\phi} ) = \frac{1}{2^{n} n! \sqrt{\pi}} e^{-x_{\phi}^{2}} H_{n} ( x_{\phi} )^{2}, 
    \end{equation}
where $H_{n}(x)$ is the Hermite polynomial of degree $n$. For the single-mode quantum states with rotationally symmetric Wigner function, e.g., Fock states, the quadrature distribution is the same for all phase angles. It is thus straightforward to obtain the value of our non-Gaussianity measure (Fig.~\ref{fig:NKL1}).

\subsubsection{Phase-averaged coherent states}
The quadrature distribution of a phase-averaged coherent state $\rho = \exp ( - \gamma^{2} ) \sum_{n=0}^{\infty} \frac{\gamma^{2}}{n!} \ketbra{n}{n}$ is given by
    \begin{equation}
        P_{\rho} ( x_{\phi} ) = \exp ( - \gamma^{2} ) \sum_{n=0}^{\infty}  \frac{\gamma^{2}}{n!} P_{\ketbra{n}{n}} ( x_{\phi} ),
    \end{equation}
with its mean photon number $\gamma^{2}$. The purity of the phase-averaged coherent state is given by
    \begin{equation}
        \mu = \exp ( -2\gamma^{2} ) I_{0} ( 2\gamma^{2} ),
    \end{equation}
where $I_{n} ( z ) $ is the modified Bessel function of the first kind of order $n$. The phase-averaged coherent state becomes non-Gaussian and mixed for nonzero $\gamma$. Also, its purity $\mu$ decreases as the coherent amplitude $\gamma$ increases. Examining the non-Gaussianity of phase-averaged coherent states, we show that our formalism is not limited to pure states (Fig.~\ref{fig:NKL1}).

\subsubsection{Cat states}
The quadrature distribution of a cat state $\ket{\psi_{\pm}} \propto \ket{\gamma} \pm \ket{-\gamma}$ is given by
    \begin{align}
        P_{\ketbra{\psi_{\pm}}{\psi_{\pm}}} ( x_{\phi} ) = & \frac{1}{\sqrt{\pi}} \frac{e^{-x_{\phi}^{2}-2\gamma^{2}\cos^{2}\phi^{2}}}{1 \pm e^{-2\gamma^{2}}} \{ \cosh ( 2\sqrt{2}\gamma x_{\phi} \cos\phi ) \nonumber \\
        & \pm \cos ( 2\sqrt{2}\gamma x_{\phi} \sin\phi ) \},
    \end{align}
with its mean photon number given by
    \begin{equation}
        \bra{\psi_{\pm}} \hat{a}^{\dag} \hat{a} \ket{\psi_{\pm}} = \gamma^{2} \frac{e^{2\gamma^{2}} \mp 1}{e^{2\gamma^{2}} \pm 1}.
    \end{equation}

In Fig.~\ref{fig:NKL1}, we plot $\mathcal{N}_{\mathrm{KL}} ( \rho )$ for Fock states and cat states with respect to the mean photon number $E = \mathrm{tr} ( \rho \hat{a}^{\dag} \hat{a} ) $. All of the states under our consideration become more non-Gaussian as the mean photon number increases. We also note that the difference in the non-Gaussianity measure between even and odd cat states becomes negligible when the mean photon number is sufficiently large. 

\subsubsection{Photon number entangled states}
The quadrature distribution of a photon number entangled state in the form of $\ket{\Psi} = \sqrt{1-f} \ket{0}_{1} \ket{0}_{2} + \sqrt{f} \ket{1}_{1} \ket{1}_{2}$ \cite{SYL2012} is given by
    \begin{align}
        P_{\ketbra{\Psi}{\Psi}} ( x_{\Theta, \Phi} ) = & \frac{e^{-x_{\Theta, \Phi}^{2}}}{8\sqrt{\pi}} \{ 8 + f ( -5 + 4 x_{\Theta, \Phi}^2 + 4 x_{\Theta, \Phi}^4 ) \nonumber \\
        & + f ( -3 + 12 x_{\Theta, \Phi}^2 - 4 x_{\Theta, \Phi}^4 ) \cos ( 4 \theta_{1} ) \nonumber \\
        & + 8 \sqrt{f(1-f)} ( 1 - 2 x_{\Theta, \Phi}^{2} ) \nonumber \\
        & \times \cos ( \phi_{1} + \phi_{2} ) \sin ( 2 \theta_{1} ) \}.
    \end{align}

\begin{figure}[!t]
		\includegraphics[scale=0.8]{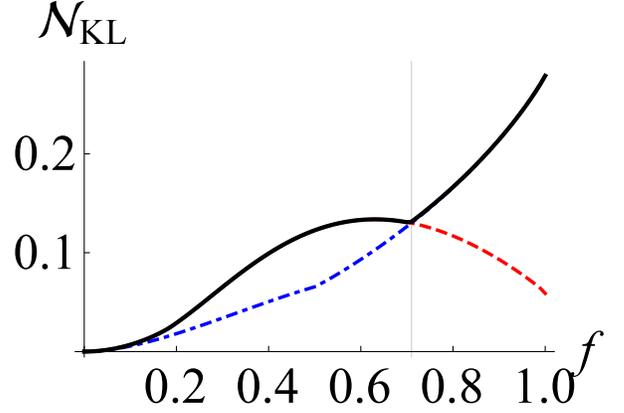}
		\caption{Our non-Gaussianity measure $\mathcal{N}_{\mathrm{KL}} ( \rho )$ for the photon number entangled state in the form of $\sqrt{1-f} \ket{0}_{1} \ket{0}_{2} + \sqrt{f} \ket{1}_{1} \ket{1}_{2}$ against the fraction $f$ (black solid line). Red dashed and blue dot-dashed lines represent the negentropies of the quadrature distributions with the maximum and minimum kurtosis, i.e., $J_{\rho, K_{\max}}$ and $J_{\rho, K_{\min}}$ in Sec. II D, respectively. The black solid line coincides with the red dashed one for $f < 0.71$ and blue dot-dashed one for $f > 0.71$.}
        \label{fig:NKL2}
	\end{figure}

In Fig.~\ref{fig:NKL2}, we show the result $\mathcal{N}_{\mathrm{KL}} ( \rho )$ for the photon number entangled state $\sqrt{1-f} \ket{0}_{1} \ket{0}_{2} + \sqrt{f} \ket{1}_{1} \ket{1}_{2}$ against the fraction $f$.

Note that a photon subtracted two-mode squeezed vacuum $\hat{a}_{1} \hat{a}_{2} \hat{S}_{12} ( \zeta ) \ket{0}_{1} \ket{0}_{2}$ can be expressed as
    \begin{align}
        & \hat{a}_{1} \hat{a}_{2} \hat{S}_{12} ( \zeta ) \ket{0}_{1} \ket{0}_{2} \nonumber \\
        & = \hat{a}_{1} \hat{S}_{12} ( \zeta ) ( - \sinh s e^{i \varphi} a_{1}^{\dag} + \cosh s \hat{a}_{2} ) \ket{0}_{1} \ket{0}_{2} \nonumber \\
        & \propto \hat{S}_{12} ( \zeta ) ( \cosh s \hat{a}_{1} - \sinh s e^{i \varphi} a_{2}^{\dag} ) \ket{1}_{1} \ket{0}_{2} \nonumber \\
        & = \hat{S}_{12} ( \zeta ) ( \cosh s \ket{0}_{1} \ket{0}_{2} - \sinh s e^{i \varphi} \ket{1}_{1} \ket{1}_{2} ) \nonumber \\
        & = \hat{S}_{12} ( \zeta ) \hat{R}_{1} ( \varphi + \pi ) ( \cosh s \ket{0}_{1} \ket{0}_{2} + \sinh s \ket{1}_{1} \ket{1}_{2} ),
    \end{align}
where we have used the transformation $\hat{S}_{12}^{\dag} ( \zeta ) \hat{a}_{1} \hat{S}_{12} ( \zeta ) = \hat{a}_{1} \cosh s - \hat{a}_{2}^{\dag} \exp ( i \varphi ) \sinh s$ and $\hat{S}_{12}^{\dag} ( \zeta ) \hat{a}_{2} \hat{S}_{12} ( \zeta ) = \hat{a}_{2} \cosh s - \hat{a}_{1}^{\dag} \exp ( i \varphi ) \sinh s$ for the two-mode squeezing operator $\hat{S}_{12} ( \zeta ) = \exp ( - \zeta \hat{a}_{1}^{\dag} \hat{a}_{2}^{\dag} + \zeta^{*} \hat{a}_{1} \hat{a}_{2} )$ with squeezing parameter $\zeta = s \exp ( i \varphi )$ \cite{BarnettRadmore}. As our non-Gaussianity measure is invariant under Gaussian unitary operation, the result in Fig.~\ref{fig:NKL2} also provides the amount of non-Gaussianity generated by the photon subtraction $\hat{a}_{1} \hat{a}_{2}$. That is, the photon number entangled state $\ket{\Psi}$ with $f = \frac{\sinh^{2} s}{\cosh (2s)}$ has the same non-Gaussianity with $\hat{a}_{1} \hat{a}_{2} \hat{S}_{12} ( \zeta ) \ket{0}_{1} \ket{0}_{2}$. Note that the non-Gaussianity of the photon-subtracted two-mode state is bounded due to $0 \leq \frac{\sinh^{2} s}{\cosh (2s)} \leq \frac{1}{2}$ for all $s$.

\subsection{Experimental feasibility}
Our non-Gaussianity measure $\mathcal{N}_{\mathrm{KL}} ( \rho )$ in Eq.~\eqref{eq:NGKL1} is defined as the maximum of the negentropy $J_{\rho} ( Q_{\Theta, \Phi} )$ over the whole set of $Q_{\Theta, \Phi}$. It may require a huge amount of experimental efforts to determine the exact value of the non-Gaussianity measure. We here propose a strategy for estimating $\mathcal{N}_{\mathrm{KL}} ( \rho )$ efficiently.

\begin{figure}[!t]
		\includegraphics[scale=0.8]{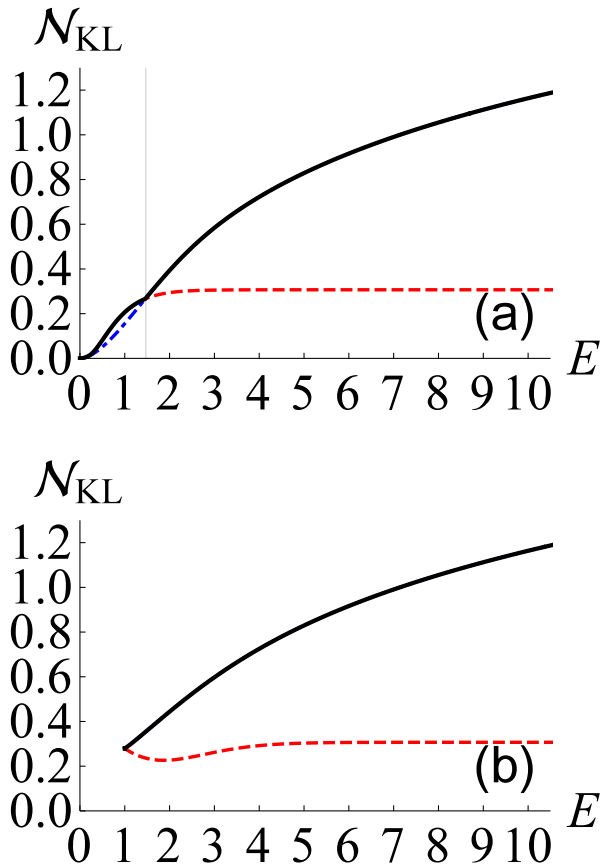}
		\caption{Non-Gaussianity measure $\mathcal{N}_{\mathrm{KL}} ( \rho )$ (black solid line) and the negentropies of the quadrature distributions with the maximum and minimum kurtosis, i.e., $J_{\rho, K_{\max}}$ (red dashed line) and $J_{\rho, K_{\min}}$ (blue dot-dashed line), for (a) even cat states and (b) odd cat states against the mean photon number $E = \mathrm{tr} ( \rho \hat{a}^{\dag} \hat{a} )$. The black solid line in (a) coincides with the red dashed and the blue dot-dashed ones for $E < 1.47$ and $E > 1.47$, respectively. In (b), the blue dot-dashed line is not discernible form the black solid line.}
        \label{fig:NKL3}
	\end{figure}

In Figs. \ref{fig:NKL2} and \ref{fig:NKL3}, we plot $\mathcal{N}_{\mathrm{KL}} ( \rho )$ and the negentropies $\mathcal{J}_{\rho} ( Q_{\Theta, \Phi} )$ of the quadrature distributions maximizing or minimizing the kurtosis $K = \frac{\expect{\Delta \hat{Q}_{\Theta, \Phi}^{4}}}{\expect{\Delta \hat{Q}_{\Theta, \Phi}^{2}}^{2}}$ \cite{Cramer1946} for the photon number entangled state and the cat states, respectively. From now on, we refer to $\mathcal{J}_{\rho, K_{\max}}$ and $\mathcal{J}_{\rho, K_{\min}}$ as the negentropy of the quadrature distribution maximizing and minimizing the kurtosis, respectively. We observe that $\mathcal{N}_{\mathrm{KL}} ( \rho )$ always coincides with one of $\mathcal{J}_{\rho, K_{\max}}$ and $\mathcal{J}_{\rho, K_{\min}}$ for the photon number entangled state and the cat states.

Our strategy goes as follows. We are to obtain the moments of {\it all} quadrature distributions up to fourth order in order to find the maximum and the minimum kurtosis. This can be done by measuring only a {\it finite} number of quadratures. For instance, the quadrature moments $\expect{\hat{q}_{\phi}^{n}}$ for an arbitrary $\phi$ can be determined by measuring the quadrature distributions fixed at $n+1$ different phase angles as $\phi_{j} = \phi_{0} + \frac{j \pi}{n+1}$ with an arbitrary $\phi_{0}$ and $j \in \{ 0, 1, ..., n \}$ \cite{Wunsche1998, Wunsche1996}. By then analyzing $\expect{\hat{q}_{\phi}^{n}}$ for all $\phi$'s, we can identify the quadratures maximizing and minimizing the kurtosis, which eventually yields the candidates for $\mathcal{N}_{\mathrm{KL}} ( \rho )$, i.e., $\mathcal{J}_{\rho, K_{\max}}$ and $\mathcal{J}_{\rho, K_{\min}}$.

\begin{figure}[!t]
		\includegraphics[scale=0.6]{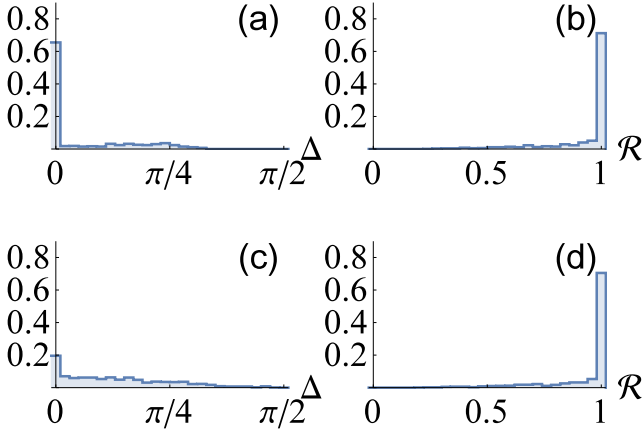}
		\caption{(a), (c): Histogram of the discrepancy $\Delta = \min [ | \phi_{\mathcal{J}_{\max}} - \phi_{K_{\max}} |, | \phi_{\mathcal{J}_{\max}} - \phi_{K_{\min}} | ]$ between the optimal parameters, i.e., the phase angles maximizing (minimizing) the kurtosis $\phi_{K_{\max}}$($\phi_{K_{\min}}$) and the negentropy $\phi_{\mathcal{J}_{\max}}$, respectively. (b), (d) The ratio $\mathcal{R} = \frac{\max [ \mathcal{J}_{\rho, K_{\max}}, \mathcal{J}_{\rho, K_{\min}} ]}{\mathcal{N}_{\mathrm{KL}} ( \rho )}$ between the estimated non-Gaussianity by the strategy described in Sec. II D, i.e., $\max [ \mathcal{J}_{\rho, K_{\max}}, \mathcal{J}_{\rho, K_{\min}} ]$ and the non-Gaussianity measure $\mathcal{N}_{\mathrm{KL}} ( \rho )$. For the histograms, we have sampled $10^{3}$ random pure states in the form of $\sum_{n=0}^{5} c_{n} \ket{n}$ with real coefficients for (a) and (b), and $10^{3}$ random mixed states in the form of $f \ketbra{\chi_{1}}{\chi_{1}} + (1-f) \ketbra{\chi_{2}}{\chi_{2}}$ for (c) and (d), where both of the pure states are in the form of $\sum_{n=0}^{5} c_{n} \ket{n}$ with real coefficients.}
        \label{fig:RPS}
	\end{figure}

To test our strategy, we examine random pure states. First we pick $10^{3}$ random pure states in the form of $\sum_{n = 0}^{5} c_{n} \ket{n}$ with real coefficients $c_{n}$. Then we investigate the discrepancy $\Delta = \min [ | \phi_{\mathcal{J}_{\max}} - \phi_{K_{\max}} |, | \phi_{\mathcal{J}_{\max}} - \phi_{K_{\min}} | ]$ between the phase angles minimizing (maximizing) the kurtosis $\phi_{K_{\min}}$($\phi_{K_{\max}}$) and maximizing the negentropy $\phi_{\mathcal{J}_{\max}}$. In Fig.~\ref{fig:RPS}(a), we plot a histogram for the discrepancy $\Delta$ with the bin size $\frac{\pi}{50}$. It shows that $\Delta$ is less than $\frac{\pi}{100}$ for $65.5 \%$ of the samples. Investigating more precisely, we observe that the optimal phase angle $\phi_{\mathcal{J}_{\max}}$ becomes $\phi_{K_{\max}}$ or $\phi_{K_{\min}}$ for $30.3 \%$ and $33.9 \%$ of the pure states, respectively. This indicates that our strategy immediately yields $\mathcal{N}_{\mathrm{KL}} ( \rho )$ for $63.9 \%$ of random pure states. In Fig.~\ref{fig:RPS}(b), we also examine the ratio $\mathcal{R}$ between $\max [ J_{\rho, K_{\max}},  J_{\rho, K_{\min}} ]$ and $\mathcal{N}_{\mathrm{KL}} ( \rho )$ to addressing the performance of our approach. We observe that the ratio becomes greater than 0.95 for $75.4 \%$ of pure states and the average ratio turns out to be 0.931. This result supports that our strategy is efficient for a wide range of quantum states.

We also extend our examination to $10^{3}$ random mixed states in the form of $f \ketbra{\chi_{1}}{\chi_{1}} + (1-f) \ketbra{\chi_{2}}{\chi_{2}}$, where both of the pure states $\ket{\chi_{1}}$ and $\ket{\chi_{2}}$ are randomly generated as a superposition of Fock states $\sum_{n=0}^{5} c_{n} \ket{n}$ with real coefficients $c_{n}$. In Figs. \ref{fig:RPS}(c) and \ref{fig:RPS}(d), we see that the histogram of the discrepancy $\Delta$ defined above is less peaked than that of Fig. \ref{fig:RPS}(a), while the histogram of the ratio $\mathcal{R}$ looks similar to Fig. \ref{fig:RPS}(b). In addition, the average ratio is obtained to be 0.928, which is very close to 0.931 for the pure states addressed in the previous paragraph. This is because the mixing of pure states smooths the phase-space distributions and reduces the fluctuation of negentropy with respect to phase angle. This result illustrates how our strategy can be robust against the mixedness of state to some extent. 

Furthermore, we address the performance of our strategy more for pure states by varying the dimension of Fock space, i.e.,  $\sum_{n = 0}^{n_{\max}} c_{n} \ket{n}$ with $n_{\max} = \{ 1, 2, 3, 4, 5 \}$. Investigating $10^{3}$ random pure states for each $n_{\max}$, we observe that the average ratio $\mathcal{R}$ is given by 0.878, 0.968, 0.966, 0.955, and 0.931 for $n_{\max} = 1$, $2$, $3$, $4$, and $5$, respectively, which suggests that our strategy may be suitable for a moderate size of dimension in Fock space. The performance can potentially be improved by coming up with more candidates than kurtosis. For instance, if we choose the phase angles maximizing and minimizing the variance as additional candidates, the average ratio increases to 0.975, 0.993, 0.987, 0.986, and 0.975 for $n_{\max} = 1$, $2$, $3$, $4$, and $5$, respectively.

\section{Estimating non-Gaussianity measure defined by quantum relative entropy}
In Secs. III and IV, we address how our non-Gaussianity measure can be related to other non-Gaussianity measures. 
In \cite{Genoni2008}, a non-Gaussianity measure of a quantum state was proposed by employing quantum relative entropy as
	\begin{equation}
		\mathcal{N}_{\mathrm{QR}} ( \rho ) \equiv S ( \rho || \rho_{\mathrm{G}} ),
	\end{equation}
where $S( \rho || \rho_{G} ) = \mathrm{tr} [ \rho ( \ln \rho - \ln \rho_{G} ) ]$ is the quantum relative entropy between $\rho$ and its reference Gaussian state $\rho_{\mathrm{G}}$ with the same first- and second-order quadrature moments as the state $\rho$. Note that we have used the subscript QR to imply that the measure is based on quantum relative entropy. Similar to the negentropy, i.e., $J ( X ) \equiv D_{\mathrm{KL}} ( X || X_{\mathrm{G}} ) = H ( X_{\mathrm{G}} ) - H ( X )$, the measure based on the quantum relative entropy can also be given by the difference between the von Neumann entropies of $\rho$ and $\rho_{G}$,
	\begin{equation}
		\mathcal{N}_{\mathrm{QR}} ( \rho ) = S_{1} ( \rho_{\mathrm{G}} ) - S_{1} ( \rho ),
	\end{equation}
where $S_{1} ( \tau ) = - \mathrm{tr} [ \tau \ln \tau ]$ is the von Neumann entropy of a quantum state $\tau$.

We here show that our measure $\mathcal{N}_{\mathrm{KL}} ( \rho )$ provides a lower bound for $\mathcal{N}_{\mathrm{QR}} ( \rho )$, i.e.,
    \begin{equation} \label{eq:QRLB}
        \mathcal{N}_{\mathrm{QR}} ( \rho ) \geq \mathcal{N}_{\mathrm{KL}} ( \rho ).
    \end{equation}

We first note that every $N$-mode state must fulfill Eq.~\eqref{eq:QRLB} if it is true for an arbitrary single-mode state. As we have stated in the previous section, the $N$-mode quadrature operator $\hat{Q}$ maximizing the negentropy can be measured by using a single-mode homodyne detection and a Gaussian unitary operation, i.e., a linear optical network, so the $N$-mode quadrature operator can be transformed to a single-mode one by the help of a Gaussian unitary operation, e.g., $\hat{Q}_{1} = \hat{U}_{\mathrm{G}} \hat{Q}_{N} \hat{U}_{\mathrm{G}}^{\dag}$, which yields $\mathcal{N}_{\mathrm{KL}} ( \rho ) = \mathcal{N}_{\mathrm{KL}} ( \mathrm{tr}_{2,...,N} [ \hat{U}_{\mathrm{G}} \rho \hat{U}_{\mathrm{G}}^{\dag} ] )$. We also note $\mathcal{N}_{\mathrm{QR}} ( \rho ) \geq \mathcal{N}_{\mathrm{QR}} ( \mathrm{tr}_{2,...,N} [ \hat{U}_{\mathrm{G}} \rho \hat{U}_{\mathrm{G}}^{\dag} ] )$, as the non-Gaussianity measure $\mathcal{N}_{\mathrm{QR}} ( \rho )$ is invariant under Gaussian unitary operation and nonincreasing under partial trace \cite{Genoni2008}.

	\begin{figure}[!t]
		\includegraphics[scale=0.8]{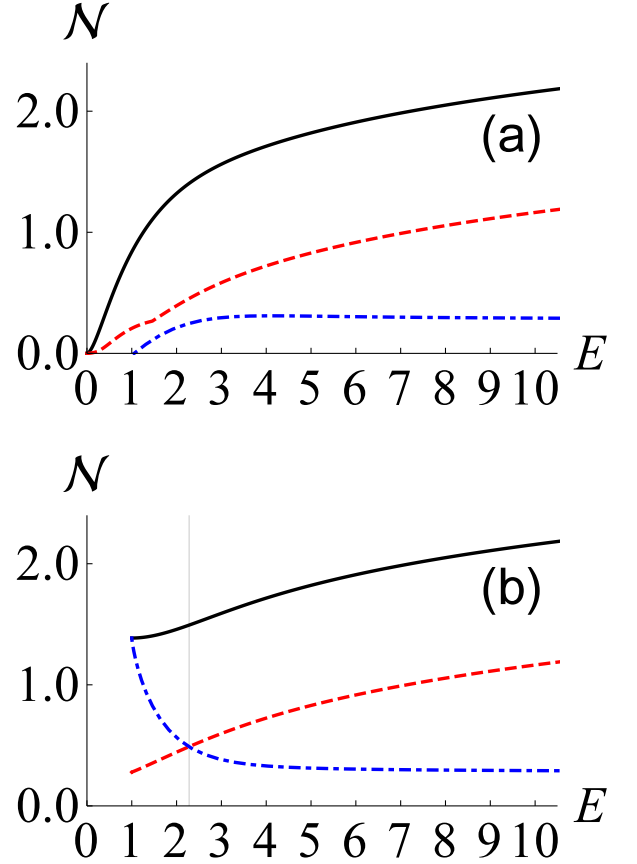}
		\caption{Non-Gaussianity measures based on quantum relative entropy $\mathcal{N}_{\mathrm{QR}} ( \rho )$ (black solid line) and Kullback-Leibler divergence $\mathcal{N}_{\mathrm{KL}} ( \rho )$ (red dashed line) and a lower bound of $\mathcal{N}_{\mathrm{QR}} ( \rho )$ in \cite{Genoni2010} (blue dot-dashed line) for (a) even cat states and (b) odd cat states with respect to the energy of the state.}
        \label{fig:LB1}
	\end{figure}

We now focus on the derivation of Eq.~\eqref{eq:QRLB} for a single-mode state $\rho$ and introduce a quantum-to-classical channel $\mathcal{T}$ \cite{Wilde2017} as
    \begin{equation}
        \rho \mapsto \mathcal{T} [ \rho ] = \sum_{n = -\infty}^{\infty} \mathrm{tr} [ \rho \hat{\Pi}_{\phi, n}^{\sigma} ] \ketbra{\psi_{n}}{\psi_{n}},
    \end{equation}
where the set of $\ket{\psi_{n}}$ forms an orthonormal basis and $\hat{\Pi}_{\phi, n}^{\sigma}$ represents a coarse-grained homodyne detection with the binning size $\sigma$ as
    \begin{equation}
        \hat{\Pi}_{\phi, n}^{\sigma} = \int_{(n-\frac{1}{2})\sigma}^{(n+\frac{1}{2})\sigma} dx_{\phi} \ketbra{x_{\phi}}{x_{\phi}},
    \end{equation}
with $\ket{x_{\phi}}$ denoting the eigenstate for a  quadrature operator $\hat{x}_{\phi}$. Using the fact that the quantum relative entropy is nonincreasing under any positive trace-preserving linear map \cite{Hermes2017}, we have
    \begin{align} \label{eq:QRLB1}
        S ( \rho || \rho_{\mathrm{G}} ) & \geq S ( \mathcal{T} [ \rho ] || \mathcal{T} [ \rho_{\mathrm{G}} ] ) \nonumber \\
        & = H ( X_{\phi, \rho}^{\sigma} || X_{\phi, \rho_{\mathrm{G}}}^{\sigma} ),
    \end{align}
where the coarse-grained quadrature distribution $X_{\phi, \rho}^{\sigma}$ is given by
        \begin{equation}
            X_{\phi, \rho}^{\sigma} ( x ) = \sum_{n = - \infty}^{\infty} \mathrm{tr}[ \rho \hat{\Pi}_{n} ] \frac{1}{\sigma} \mathrm{rect} [ \frac{x}{\sigma} - n ]
    \end{equation}
with
    \begin{equation}
        \mathrm{rect} [ x ] =
        \begin{cases}
            1 & \mbox{for $|x| \leq \frac{1}{2}$}, \\
            0 & \mbox{for $|x| > \frac{1}{2}$}.
        \end{cases}
    \end{equation}
Note that
    \begin{equation}
        H ( X_{\phi, \rho} || X_{\phi, \rho_{\mathrm{G}}} ) = \sum_{n = - \infty}^{\infty} \mathrm{tr} [ \rho \Pi_{n} ] \ln \frac{\mathrm{tr} [ \rho \Pi_{n} ]}{\mathrm{tr} [ \rho_{\mathrm{G}} \Pi_{n} ]}.
    \end{equation}
Using the log sum inequality \cite{CoverThomas}, i.e.,
    \begin{equation}
        \sum_{k} a_{k} \ln \frac{a_{k}}{b_{k}} \geq a \ln \frac{a}{b},
    \end{equation}
with $a = \sum_{k} a_{k}$ and $b = \sum_{k} b_{k}$, we observe that
    \begin{equation}
        H ( X_{\phi, \rho}^{\sigma} || X_{\phi, \rho_{\mathrm{G}}}^{\sigma} ) \leq H ( X_{\phi, \rho}^{\frac{\sigma}{M}} || X_{\phi, \rho_{\mathrm{G}}}^{\frac{\sigma}{M}} ),
    \end{equation}
for any positive integer $M$. It indicates that the relative entropy increases with the decrease of coarse-graining size,  so
    \begin{align} \label{eq:QRLB2}
        \sup_{\sigma} H ( X_{\phi, \rho}^{\sigma} || X_{\phi, \rho_{\mathrm{G}}}^{\sigma} ) & = \lim_{\sigma \rightarrow 0} H ( X_{\phi, \rho}^{\sigma} || X_{\phi, \rho_{\mathrm{G}}}^{\sigma} ) \nonumber \\
        & = H ( X_{\phi, \rho} || X_{\phi, \rho_{\mathrm{G}}} ).
    \end{align}
Combining Eqs. \eqref{eq:NGKL2}, \eqref{eq:QRLB1}, and \eqref{eq:QRLB2}, we finally obtain
    \begin{align}
        S ( \rho || \rho_{\mathrm{G}} ) & \geq \max_{\phi} \sup_{\sigma} H ( X_{\phi, \rho}^{\sigma} || X_{\phi, \rho_{\mathrm{G}}}^{\sigma} ) \nonumber \\
        & = \max_{\phi} H ( X_{\phi, \rho} || X_{\phi, \rho_{\mathrm{G}}} ) \nonumber \\
        & = \mathcal{N}_{\mathrm{KL}} ( \rho ).
    \end{align}

In Fig. \ref{fig:LB1} we plot $\mathcal{N}_{\mathrm{QR}} ( \rho )$, $\mathcal{N}_{\mathrm{KL}} ( \rho )$ and a lower bound of $\mathcal{N}_{\mathrm{QR}} ( \rho )$ proposed in \cite{Genoni2010} for even and odd cat states. The lower bound in \cite{Genoni2010} can be obtained by measuring covariance matrix and photon number distribution, i.e., $\mathcal{N}_{\mathrm{QR}} ( \rho ) \geq \mathcal{L} ( \rho ) \equiv S ( \rho_{\mathrm{G}} ) - \sum_{k=0} P_{n} \ln P_{n}$ with $P_{n} = \bra{n} \rho \ket{n}$. We find that $\mathcal{N}_{\mathrm{KL}} ( \rho )$ provides a greater lower bound for $\mathcal{N}_{\mathrm{QR}} ( \rho )$ than $\mathcal{L} ( \rho )$ for all even cat states and odd cat states with $E > 2.28$ ($\gamma > 1.49$). The result shows that our non-Gaussianity measure $\mathcal{N}_{\mathrm{KL}} ( \rho )$ provides an efficient tool to estimate $\mathcal{N}_{\mathrm{QR}} ( \rho )$ without quantum-state tomography, especially for quantum states without rotational symmetry in phase space. Note that the odd cat state approaches a single-photon state, i.e., a quantum state with rotational symmetry in phase space, as the coherent amplitude $\gamma$ decreases.

\subsection{Application in entanglement detection}
It is worth noting that the inequality \eqref{eq:QRLB} can be used to derive a new uncertainty relation whose bound is determined by the non-Gaussianity and the entropy of the state  (cf. \cite{Baek2018}). For a Gaussian state, we have the identity $S_{1} ( \rho_{\mathrm{G}} ) = h ( \sqrt{\det \Gamma} )$, where $h(x)=(x+\frac{1}{2})\ln(x+\frac{1}{2})-(x-\frac{1}{2})\ln (x-\frac{1}{2})$. That is, the von Neumann entropy of a Gaussian state is completely determined by the covariance matrix. The inequality \eqref{eq:QRLB}, which can be written as 
$S_{1} ( \rho_{\mathrm{G}} ) \geq S_{1} ( \rho)+\mathcal{N}_{\mathrm{KL}} (\rho)$, then leads to $h(\sqrt{\det \Gamma(\rho)})\geq \mathcal{N}_{\mathrm{KL}} (\rho)+S_1(\rho)$. Therefore, we obtain
    \begin{align} \label{eq:UR}
        \sqrt{\det \Gamma(\rho)} \geq h^{-1}( \mathcal{N}_{\mathrm{KL}} (\rho)+ S_{1}(\rho)),
    \end{align}
where $h^{-1}(y)$ is the inverse of the monotonically increasing function $h(x)$. This uncertainty relation can be considered as a generalization of the Robertson-Schr\"{o}dinger (RS) uncertainty relation $\sqrt{\det \Gamma} \geq \frac{1}{2}$, because the relation \eqref{eq:UR} gives a stronger bound, particularly for non-Gaussian or mixed states, i.e., when $\mathcal{N}_{\mathrm{KL}} ( \rho ) > 0$ or $S_{1} ( \rho ) > 0$, respectively.

Non-Gaussianity- and entropy-bounded uncertainty relations such as Eq. \eqref{eq:UR} are potentially applicable to improve Simon-Duan entanglement criterion \cite{Simon2000,Duan2000}, which is a necessary and sufficient criterion for Gaussian states only. For example, if the inequality Eq. \eqref{eq:UR} is violated under partial transposition, it is a direct signature of quantum entanglement. The inequality, Eq. \eqref{eq:UR}, thus leads to improved entanglement criteria, particularly for non-Gaussian entangled states, like those in \cite{Baek2018}.

Let us explain how it works in detail. We first follow the procedure of positive partial transposition (PPT) criterion for Gaussian entanglement. The Gaussian PPT criterion identifies a quantum state $\rho$ as entangled if the covariance matrix becomes unphysical under partial transposition (PT). Checking whether the covariance matrix of the partially transposed quantum state $\rho^{\mathrm{PT}}$ is physical or not, we use a symplectic transformation $\hat{S}$ which diagonalizes the covariance matrix of $\rho^{\mathrm{PT}}$. Because of the Williamson theorem \cite{Williamson1936}, such symplectic transformation always exists. If there exists a local mode violating the RS uncertainty relation $\sqrt{\det \Gamma} \geq \frac{1}{2}$, it signifies that $\rho^{\mathrm{PT}}$ is unphysical and the Gaussian PPT criterion detects the entanglement of $\rho$. In summary, the Gaussian PPT criterion can be written as
    \begin{equation}
        \min_{i} \sqrt{\det \Gamma ( \overline{\rho}_{i} )} \geq \frac{1}{2},
    \end{equation}
where $\overline{\rho}_i$ denotes the $i$th local mode of $\overline{\rho} = \hat{S} \rho^{\mathrm{PT}} \hat{S}^{\dag}$. We here employ Eq.~\eqref{eq:UR} instead of the RS uncertainty relation:
    \begin{equation} \label{eq:NGEC}
        \min_{i} \sqrt{\det \Gamma ( \overline{\rho}_{i} )} \geq h^{-1} ( \mathcal{N}_{\mathrm{KL}} ( \overline{\rho}_{i} ) ) \geq \frac{1}{2},
    \end{equation}
which is strictly stronger than the Gaussian PPT criterion. One may doubt that $\mathcal{N}_{\mathrm{KL}} ( \overline{\rho}_{i} )$ is experimentally accessible because PT cannot be directly implemented. We stress that it is possible to estimate $\mathcal{N}_{\mathrm{KL}} ( \overline{\rho}_{i} )$, as a multimode quadrature operator transforms to another one under partial transposition always.

	\begin{figure}[!t]
		\includegraphics[scale=0.8]{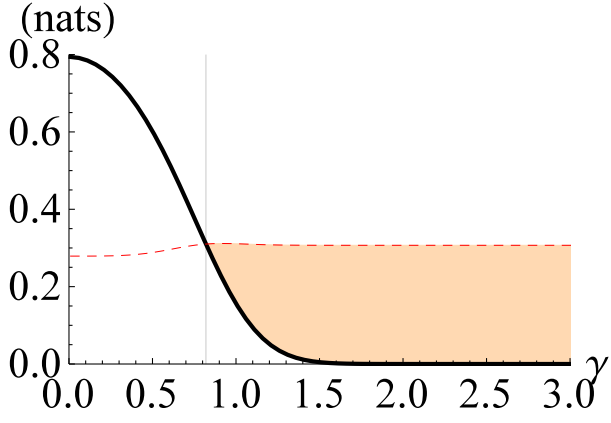}
		\caption{The entropic quantities $h ( \sqrt{\det \Gamma ( \overline{\rho}_{2} )} )$ and $\mathcal{N}_{\mathrm{KL}} ( \overline{\rho}_{2} )$ for an entangled coherent state $\ket{\Psi} = \sqrt{\mathscr{N}} ( \ket{\gamma}_{1} \ket{\gamma}_{2} - \ket{- \gamma}_{1} \ket{- \gamma}_{2} )$ are plotted as black thick and red dashed curves, respectively, with respect to the coherent amplitude $\gamma$. As Eq.~\eqref{eq:NGEC} is satisfied for every separable quantum state, its violation, i.e., $h ( \sqrt{\det \Gamma ( \overline{\rho}_{2} )} ) < \mathcal{N}_{\mathrm{KL}} ( \overline{\rho}_{2} )$, witnesses quantum entanglement. The shaded region indicates that Eq.~\eqref{eq:NGEC} can detect the quantum entanglement of $\ket{\Psi}$ with $\gamma > 0.82$ that is impossible to detect by Gaussian PPT criterion. Note that $\ln 2$ nats is equivalent to 1 bit.}
        \label{fig:ECS}
	\end{figure}

For a concrete example, we here examine an entangled coherent state $\ket{\Psi} = \sqrt{\mathscr{N}} ( \ket{\gamma}_{1} \ket{\gamma}_{2} - \ket{-\gamma}_{1} \ket{-\gamma}_{2} )$ with $\mathscr{N} = \{ 2 - 2\exp(-4\gamma^{2}) \}^{-1}$, whose entanglement cannot be detectable by the Gaussian PPT criterion for all nonzero $\gamma$. If we apply partial transposition on the second mode of $\rho = \ketbra{\Psi}{\Psi}$, the partially transposed quantum state $\rho^{\mathrm{PT}}$ is expressed by $\rho^{\mathrm{PT}} = \mathscr{N} ( \ketbra{\gamma}{\gamma}_{1} \otimes \ketbra{\gamma}{\gamma}_{2} + \ketbra{-\gamma}{-\gamma}_{1} \otimes \ketbra{-\gamma}{-\gamma}_{2} - \ketbra{\gamma}{-\gamma}_{1} \otimes \ketbra{-\gamma}{\gamma}_{2} - \ketbra{-\gamma}{\gamma}_{1} \otimes \ketbra{\gamma}{-\gamma}_{2} )$. Its covariance matrix can be diagonalized by a 50:50 beam-splitting operation $\hat{U}_{\mathrm{BS}}$. If we look into the local states of $\overline{\rho} \equiv \hat{U}_{\mathrm{BS}} \rho^{\mathrm{PT}} \hat{U}_{\mathrm{BS}}^{\dag} = \mathscr{N} \{ ( \ketbra{\sqrt{2}\gamma}{\sqrt{2}\gamma}_{1} + \ketbra{-\sqrt{2}\gamma}{-\sqrt{2}\gamma}_{1} ) \otimes \ketbra{0}{0}_{2} + \ketbra{0}{0}_{1} \otimes ( \ketbra{\sqrt{2}\gamma}{-\sqrt{2}\gamma}_{2} + \ketbra{-\sqrt{2}\gamma}{\sqrt{2}\gamma}_{2} ) \}$, we find that the local state for the mode 1 is physical, i.e., $\overline{\rho}_{1} = \mathscr{N} \{ \ketbra{\sqrt{2}\gamma}{\sqrt{2}\gamma} + \ketbra{-\sqrt{2}\gamma}{-\sqrt{2}\gamma} + 2 \exp (-4\gamma^{2}) \ketbra{0}{0} \}$, but the one for mode 2 is unphysical, i.e., $\overline{\rho}_{2} = \mathscr{N} ( 2 \ketbra{0}{0} - \ketbra{\sqrt{2}\gamma}{-\sqrt{2}\gamma} - \ketbra{-\sqrt{2}\gamma}{\sqrt{2}\gamma} )$. In Fig. \ref{fig:ECS} we compare $h ( \sqrt{\det \Gamma ( \overline{\rho}_{2} )} )$ and $\mathcal{N}_{\mathrm{KL}} ( \overline{\rho_{2}} )$ to check whether Eq.~\eqref{eq:NGEC} is violated or not. The result shows that our entanglement criterion can detect the entanglement of $\ket{\Psi}$ for $\gamma > 0.82$.

\section{Estimation of non-Gaussianity measure by Hilbert-Schmidt distance}
In \cite{Genoni2007}, a non-Gaussianity measure of a quantum state was proposed by using Hilbert-Schmidt distance as
	\begin{equation}
		\mathcal{N}_{\mathrm{HS}} ( \rho ) = \frac{D_{\mathrm{HS}} ( \rho , \rho_{\mathrm{G}} )}{2 \mathrm{tr} \rho^{2}},
	\end{equation}
where $D_{\mathrm{HS}} ( \rho, \rho_{\mathrm{G}} ) = \mathrm{tr} ( \rho - \rho_{\mathrm{G}} )^{2}$ is the Hilbert-Schmidt distance between $\rho$ and $\rho_{\mathrm{G}}$.

	\begin{figure}[!t]
		\includegraphics[scale=0.8]{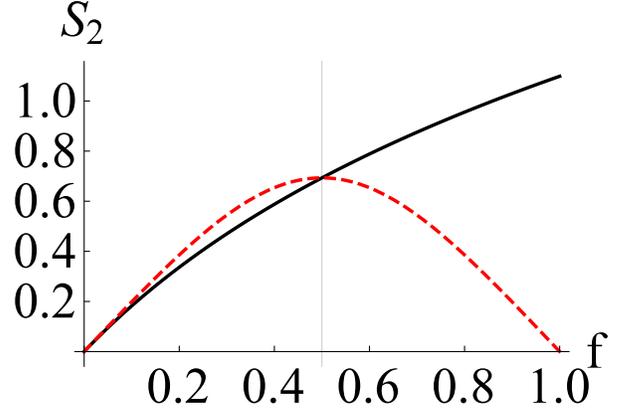}
		\caption{Quantum R{\'e}nyi-2 entropy for noisy single-photon state $(1-f) \ketbra{0}{0} + f \ketbra{1}{1}$ (red dashed line) and its reference Gaussian state (black solid line). It is clear that quantum R{\'e}nyi-2 entropy has no Gaussian extremality.}
	    \label{fig:QR2}
	\end{figure}

Using a Cauchy-Schwarz inequality, we first have
	\begin{align} \label{eq:CS}
		\mathrm{tr} ( AB ) = & \sum_{i} \sum_{j} A_{ij} B_{ji} \nonumber \\
		\leq & \sqrt{\sum_{i} \sum_{j} A_{ij} A_{ij}^{*}} \sqrt{\sum_{i} \sum_{j} B_{ji} B_{ji}^{*}} \nonumber \\
		= & \sqrt{\sum_{i} \sum_{j} A_{ij} A_{ji}} \sqrt{\sum_{i} \sum_{j} B_{ji} B_{ij}} \nonumber \\
		= & \sqrt{\mathrm{tr} A^{2}} \sqrt{\mathrm{tr} B^{2}},
	\end{align}
for two Hermitian matrices $A$ and $B$. Employing this inequality, we obtain
	\begin{align}
		\mathcal{N}_{\mathrm{HS}} ( \rho ) & = \frac{1}{2} \frac{\mathrm{tr} \rho^{2} + \mathrm{tr} \rho_{\mathrm{G}}^{2} - 2 \mathrm{tr} ( \rho \rho_{\mathrm{G}} )}{\mathrm{tr} \rho^{2}} \nonumber \\
		& \geq \frac{1}{2} \frac{\mathrm{tr} \rho^{2} + \mathrm{tr} \rho_{\mathrm{G}}^{2} - 2 \sqrt{\mathrm{tr} \rho^{2}} \sqrt{\mathrm{tr} \rho_{\mathrm{G}}^{2}}}{\mathrm{tr} \rho^{2}} \nonumber \\
		& = \frac{1}{2} \bigg( 1 - \sqrt{\frac{\mathrm{tr} \rho_{\mathrm{G}}^{2}}{\mathrm{tr} \rho^{2}}} \bigg)^{2} \nonumber \\
		& = \frac{1}{2} \bigg\{ 1 - \exp \bigg( \frac{S_{2} ( \rho ) - S_{2} ( \rho_{\mathrm{G}} )}{2} \bigg) \bigg\}^{2},
	\end{align}
where the quantum R{\'e}nyi-2 entropy of a quantum state $\rho$ is given by $S_{2} ( \rho ) = - \ln \mathrm{tr} \rho^{2}$. For an $N$-mode Gaussian state $\sigma$, we have 
    \begin{equation} \label{eq:D0}
        S_{2} ( \sigma ) - S_{1} ( \sigma ) \geq N \ln \frac{2}{e},
    \end{equation}
with its proof given in the Appendix. Due to the ordering relation of quantum R{\'e}nyi entropies, $S_{1} ( \rho ) \geq S_{2} ( \rho )$, with Eq. \eqref{eq:D0}, we obtain
    \begin{align} \label{eq:DLB}
       S_{2} ( \rho_{\mathrm{G}} ) - S_{2} ( \rho ) & \geq S_{1} ( \rho_{\mathrm{G}} ) + N \ln \frac{2}{e} - S_{2} ( \rho ) \nonumber \\ 
       & \geq S_{1} ( \rho_{\mathrm{G}} ) + N \ln \frac{2}{e} - S_{1} ( \rho ) \nonumber \\
       & \geq \mathcal{N}_{\mathrm{QR}} ( \rho ) + N \ln \frac{2}{e}.
    \end{align}
Using Eqs. \eqref{eq:QRLB} and \eqref{eq:DLB}, we obtain
    \begin{equation} \label{eq:HSLB}
        \mathcal{N}_{\mathrm{HS}} ( \rho ) \geq \frac{1}{2} \{ 1 - \mathcal{F}_{N} ( \rho ) \}^{2},
    \end{equation}
with
    \begin{equation}
        \mathcal{F}_{N} ( \rho ) = \min \bigg[ 1, \exp \bigg\{ - \frac{\mathcal{N}_{\mathrm{KL}} ( \rho )}{2} + \frac{N}{2} \ln \frac{e}{2} \bigg\} \bigg].
    \end{equation}
Note that $S_{2} ( \rho_{\mathrm{G}} ) - S_{2} ( \rho )$ can be negative while $\mathcal{N}_{\mathrm{QR}} ( \rho ) = S_{1} ( \rho_{\mathrm{G}} ) - S_{1} ( \rho )$ is always non-negative. It means that the quantum R{\'e}nyi-2 entropy has no Gaussian extremality \cite{Wolf2006}. Furthermore, Eq. \eqref{eq:DLB} reveals that $S_{2} ( \rho_{\mathrm{G}} ) - S_{2} ( \rho ) <0$ happens only if the non-Gaussianity of the quantum state $\rho$ is sufficiently small, i.e., $\mathcal{N}_{\mathrm{KL}} ( \rho ) \leq \mathcal{N}_{\mathrm{QR}} ( \rho ) \leq N \ln \frac{e}{2}$. For example, a noisy single-photon state in the form of $f \ketbra{1}{1} + (1-f) \ketbra{0}{0}$ shows $S_{2} ( \rho_{\mathrm{G}} ) < S_{2} ( \rho )$ for $0 < f < \frac{1}{2}$, as plotted in Fig. \ref{fig:QR2}.

We remark that Eq.~\eqref{eq:DLB} also yields a relation between $\mathcal{N}_{\mathrm{QR}} ( \rho )$ and the trace overlap $\mathcal{O} ( \rho ) = \mathrm{tr} ( \rho \rho_{\mathrm{G}} )$ between a state $\rho$ and its reference Gaussian state $\rho_{\mathrm{G}}$ \cite{Mandilara2010} as
    \begin{equation} \label{eq:OL}
        \frac{\mathcal{O} ( \rho )}{\mu ( \rho )} \leq \bigg( \frac{e}{2} \bigg)^{\frac{N}{2}} \exp \bigg\{ - \frac{\mathcal{N}_{\mathrm{QR}} ( \rho )}{2} \bigg\},
    \end{equation}
which implies that $\mathcal{N}_{\mathrm{QR}} ( \rho )$ provides an upper bound for the ratio of the overlap $\mathcal{O} ( \rho )$ to the purity $\mu ( \rho ) = \mathrm{tr} \rho^{2}$. The overlap $\mathcal{O} ( \rho )$ becomes identical to the purity $\mu ( \rho )$ for $\rho = \rho_{\mathrm{G}}$, and the deviation between the overlap $\mathcal{O} ( \rho )$ and the purity $\mu ( \rho )$ can be seen as a degree of non-Gaussianity \cite{Mandilara2010}. For instance, if a quantum state $\rho$ satisfies $\mathcal{O} ( \rho ) \ll \mu ( \rho )$, it witnesses that the quantum state $\rho$ is highly non-Gaussian in terms of the ratio $\frac{\mathcal{O} ( \rho )}{\mu ( \rho )}$. Therefore Eq.~\eqref{eq:OL} indicates that a larger $\mathcal{N}_{\mathrm{QR}} ( \rho )$ guarantees a higher non-Gaussianity in terms of the ratio, i.e., a smaller upper bound of $\frac{\mathcal{O} ( \rho )}{\mu ( \rho )}$. Note that we can derive Eq.~\eqref{eq:OL} by using the Cauchy-Schwarz inequality in Eq.~\eqref{eq:CS}, i.e., $\mathcal{O} ( \rho ) \leq \sqrt{\mu ( \rho ) \mu ( \rho_{\mathrm{G}} )}$, and Eq.~\eqref{eq:DLB}, i.e., $- \ln \frac{\mu ( \rho_{\mathrm{G}} )}{\mu ( \rho )} \geq \mathcal{N}_{\mathrm{QR}} ( \rho ) + N \ln \frac{2}{e}$ as
    \begin{align}
        \frac{\mathcal{O} ( \rho )}{\mu ( \rho )} & \leq \sqrt{\frac{\mu ( \rho_{\mathrm{G}} )}{\mu ( \rho )}} \nonumber \\
        & \leq \exp \bigg\{ - \frac{\mathcal{N}_{\mathrm{QR}} ( \rho )}{2} - \frac{N}{2} \ln \frac{2}{e} \bigg\} \nonumber \\
        & = \bigg( \frac{e}{2} \bigg)^{\frac{N}{2}} \exp \bigg\{ - \frac{\mathcal{N}_{\mathrm{QR}} ( \rho )}{2} \bigg\}.
    \end{align}

\section{Concluding Remarks}
We have proposed the maximum negentropy of quadrature distributions as a non-Gaussianity measure of a general $N$-mode quantum state. Our measure fulfills desirable properties, i.e., it is faithful, invariant under a Gaussian unitary operation, and nonincreasing under a trace-preserving Gaussian channel.  Furthermore, we have shown that our measure provides lower bounds for other non-Gaussianity measures based on quantum relative entropy and Hilbert-Schmidt distance, respectively. As our measure is experimentally accessible by a highly efficient homodyne detection, the connection between our measure and others makes it possible to address the issue of non-Gaussianity in an experimentally friendly form. Therefore we hope our approach could be broadly adopted in assessing the role of non-Gaussianity in continuous-variable quantum information protocols.

Recently, quantum non-Gaussianity, i.e., a stronger form of non-Gaussianity, has attracted much attention in continuous-variable quantum information. It is because non-Gaussian states generated by mixing Gaussian states can fail to be genuine quantum resources in various quantum tasks \cite{Albarelli2018, Takagi2018, Lloyd1999, Menicucci2006, Lachman2021}. For instance, a quantum circuit composed of quantum states, operations, and measurements with positive Wigner functions cannot show quantum advantage as it is classically simulable \cite{Mari2012}. It will be truly significant if one finds an efficient way to assess the more robust forms of non-Gaussianity, i.e., quantum non-Gaussianity and negativity in Wigner phase space. In addition, it would be interesting to further extend our approach to quantify the non-Gaussianity of correlation \cite{Park2017}, which we leave as future work.

\section*{Acknowledgments}
J.P. acknowledges support by the National Research Foundation of Korea (NRF) grant funded by the Korea government (MSIT) (NRF-2019R1G1A1002337). J.L. and K.B. supported by KIAS Individual Grants (No. CG073102 and No. CG074702) at Korea Institute for Advanced Study, respectively. K.B was supported by an NRF Grant funded by the Government of Korea (MSIT) (No. 2020M3E4A1079939).

\bibliographystyle{apsrev}

\section*{Appendix}
An $N$-mode Gaussian state $\sigma$ can be transformed to an $N$-mode thermal state by a Gaussian unitary operation $\hat{U}_{S}$ \cite{Weedbrook2012}:
    \begin{equation}
        \hat{U}_{S} \sigma \hat{U}_{S}^{\dag} = \tau_{\bar{n}_{1}} \otimes \tau_{\bar{n}_{2}} \otimes \cdots \otimes\tau_{\bar{n}_{N}},
    \end{equation}
where $\tau_{\bar{n}} = \sum_{k=0}^{\infty} \frac{\bar{n}^{k}}{(\bar{n}+1)^{k+1}} \ketbra{k}{k}$ represents the thermal state with the mean photon number $\bar{n}$. As quantum R{\'e}nyi entropies are invariant under unitary operations and additive for product states, i.e., $S_{\alpha} ( \rho ) = S_{\alpha} ( \hat{U} \rho \hat{U}^{\dag} )$ and $S_{\alpha} ( \rho_{A} \otimes \rho_{B} ) = S_{\alpha} ( \rho_{A} ) + S_{\alpha} ( \rho_{B} )$, respectively, we have
	\begin{align} \label{eq:D1}
		S_{2} ( \sigma ) - S_{1} ( \sigma ) & = S_{2} ( \hat{U}_{S} \sigma \hat{U}_{S}^{\dag} ) - S_{1} ( \hat{U}_{S} \sigma \hat{U}_{S}^{\dag} ) \nonumber \\
		& = \sum_{j=1}^{N} \{ S_{2} ( \tau_{\bar{n}_{j}} ) - S_{1} ( \tau_{\bar{n}_{j}} ) \},
	\end{align}
with $S_{1} ( \tau_{\bar{n}} ) = ( \bar{n} + 1 ) \ln ( \bar{n} + 1 ) - \bar{n} \ln \bar{n}$ and $ S_{2} ( \tau_{\bar{n}} ) = \ln ( 1 + 2 \bar{n} )$. Investigating the first and second derivatives of $\mathcal{D} ( \bar{n} ) \equiv S_{1} ( \tau_{\bar{n}} ) - S_{2} ( \tau_{\bar{n}} )$,
	\begin{align}
		\lim_{\bar{n} \rightarrow \infty} \frac{d}{d \bar{n}} \mathcal{D} ( \bar{n} ) = & \lim_{\bar{n} \rightarrow \infty} \bigg[ \ln \frac{1 + \bar{n}}{\bar{n}} - \frac{2}{1 + 2 \bar{n}} \bigg] = 0,
	\end{align}
and
	\begin{equation}
		\frac{d^{2}}{d\bar{n}^{2}} \mathcal{D} ( \bar{n} ) = - \frac{1}{\bar{n} ( 1 + \bar{n} ) ( 1 + 2 \bar{n} )^{2}},
	\end{equation}
we find that the difference is monotonically increasing with respect to $\bar{n}$. It yields
    \begin{align} \label{eq:D2}
        \mathcal{D} ( \bar{n} ) & \leq \lim_{\bar{n} \rightarrow \infty} \mathcal{D} ( \bar{n} ) \nonumber \\
        & = \lim_{\bar{n} \rightarrow \infty} \bigg[ \ln \bigg( 1 + \frac{1}{\bar{n}} \bigg)^{\bar{n}} + \ln \frac{1 + \bar{n}}{1 + 2 \bar{n}} \bigg] \nonumber \\
        & = \ln \frac{e}{2}.
    \end{align}
Using Eqs. \eqref{eq:D1} and \eqref{eq:D2}, we have
    \begin{equation}
        S_{2} ( \sigma ) \geq S_{1} ( \sigma ) + N \ln \frac{2}{e}.
    \end{equation}
\end{document}